# Seeing Through Misinformation:
# A Framework for Identifying Fake Online News

by Murphy Choy, DPS[1] and Mark Chong, PhD[2]

**Abstract**

The fake news epidemic makes it imperative to develop a diagnostic framework that is both parsimonious and valid to guide present and future efforts in fake news detection. This paper represents one of the very first attempts to fill a void in the research on this topic. The LeSiE (Lexical Structure, Simplicity, Emotion) framework we created and validated allows lay people to identify potential fake news without the use of calculators or complex statistics by looking out for three simple cues.

**Introduction**

A panel of experts convened by the BBC in 2016 named the breakdown of trusted sources of information as one of the most pressing societal challenges in the 21st century. In the same year, the Oxford Dictionaries named "post-truth" the "word of the year". The outcomes of two of the most momentous events in 2016 — the US presidential election and Brexit — were thought to have been significantly influenced by the prevalence of fake news surrounding both events.

The speed with which misinformation make its way online and find an audience is unprecedented in the history of communication. This phenomenon is fuelled by a combination of the ubiquity of social networks, credulous online media, and the peculiarities of human information processing (Silverman, 2015). As fake news articles are unconstrained by reality or facts, they can be crafted with considerable latitude to appeal to hopes, fears, wishes and curiosity, which in turn drives online virality and engagement (Silverman, 2015). This is complicated by the popularity of rumors as conversational currency in interpersonal interactions, especially under conditions of uncertainty (e.g. Southwell, 2013).

Fake news viewing has been shown to foster feelings of inefficacy, alienation, and cynicism (Balmas, 2014). Moreover, fakes news (if plausible) can create a complex and demanding rhetorical situation for organizations (Veil et al., 2012), such that organizational legitimacy could be threatened and undermined (Sellnow, Littlefield, Vidolof, & Webb, 2009). On the political/international stage, the consequences of fake news can take on epic proportions: for example, over half of those who recalled seeing fake news about Donald Trump and Hilary Clinton during the US presidential election actually believed them (Allcott & Gentzkow, 2017). In a separate study, Balmas (2014) showed that strong exposure to fake news was associated with higher perceived realism of the fake news.

---

[1] Director of Operations and Technology, SSON Analytics.
[2] Associate Professor, Corporate Communication (Practice), Singapore Management University.

In June 2017, the Asia News Network – a regional alliance of 22 media organizations across Asia – pledged to work together to develop a checklist that will help reporters and readers spot tell-tale signs of fake news. Other news, civic, and governmental organizations around the world are doing the same. Given this dire backdrop, there is an urgent need to develop a diagnostic framework that is both parsimonious and valid to guide present and future efforts in fake news detection. This study represents one of the very first attempts to fill the void in the literature on this topic.

**Literature Review**

Fake news comprises information that is intentionally created for media distribution to create a false impression or conclusion (Burgoon, Buller, Guerrero, Afifi, & Feldman, 1996). Just as liars must carefully manipulate language to create a convincing story and present it in a manner that is perceived as truthful by others (Friedman & Tucker, 1990), creators of fake news must do the same to engender believability. Fake news and other forms of deception involve the careful construction of messages or stories and the manipulation of language to give an impression of truthfulness (Zhou & Zhang, 2008). Therefore, stories or messages based on manufactured experiences that have little or no basis in reality are qualitatively different from stories that are anchored in real experience (e.g. Johnson & Raye, 1981; Vrij, Edward, Roberts, & Bull, 2000). By looking at the type of language used in news stories, researchers can tell the differences between true and false stories (Newman, Pennebaker, Berry, & Richards, 2003).

A growing body of research has shown that people's underlying thoughts, emotions, and motives can be revealed through the specific words they use in their communication – linguistic cues are rooted in the psychological experience underlying deception (see Zhou & Zhang, 2008). In particular, words that reflect how people express themselves can often be more revealing about their underlying thoughts, emotions, and motives than what they express (Pennebaker & King, 1999; Pennebaker, Mehl, & Niederhoffer, 2003). While liars may have some control over information presented in their stories, "their underlying state of mind may leak out through the style of language used to tell the story" (Newman et al., 2003, p. 672). Accordingly, these linguistic 'slips of tongue' can reveal underlying anxiety, guilt, or arousal (Zhou & Zhang, 2008). Pronoun use, emotionally toned words, prepositions and conjunctions are some of the features of linguistic use that have been linked to a number of behavioural and emotional outcomes associated with deception (Newman et al., 2003).

Detecting fake news is particularly challenging for two reasons. First, the nature of the news media largely precludes any observation of the non-verbal behaviour of the sender of the fake news (see Zhou & Zhang, 2008). Second, creators of fake news likely have more time and opportunities to craft their mediated messages and stories than in face-to-face communication — this reduces the 'leakage' of deception behaviour (see Zhou, Burgoon, Nunamaker, & Twitchell, 2004). There are also a number of human psychological biases that complicate fake news detection – these include belief perseverance (see DiFonzo & Bordia, 2007), the "backfire effect" (see Nyhan & Reifler, 2010), confirmation bias (see Nickerson, 1998), biased assimilation (see Lord, Ross, & Lepper, 1979), group polarization (see Boyd & Yardi, 2010), and denial transparency (see Coulton, Wegner, & Wenzlaff, 1985).

There have been several attempts by journalistic (IFLA, 2017; "Associated Press," 2017) and community organizations ("Tips to Spot Fake News," 2017; Einstein, 2017) to develop frameworks or guides to identify fake news. However, many of the frameworks have more than four steps and involve many layers of checks that will be impossible for any individual to perform without access to the Internet or, in some cases, access to reference-checking resources. Thus, it is critical to have a simpler and easier framework to identify fake news to prevent its spread and malevolent impact.

We propose a framework to identify fake news that simultaneously considers three dimensions which feature independently in other frameworks – lexical structure, simplicity, and emotion:

1. Lexical Structure

Fake news has lexical structures[3] that are different from factual reports (Rubin, Conroy, Chen, & Cornwall, 2016; Berghel, 2017). For example, biased information has been associated with specific linguistic cues, including active verbs, implicatives, hedges, and subjective intensifiers (Recasens, Danescu-Niculescu-Mizil, & Jurafsky, 2013). 'Clickbait' — a form of misleading online content — has unique lexical features such as unresolved pronouns, affective language, action words, suspenseful language, and the overuse of numerals (Chen, Conroy, & Rubin, 2015). Since fake news has unique lexical features, we can use these features to detect its presence in online news content.

2. Simplicity

Deceivers (e.g. creators of fake news) tend to tell less complex stories (Newman et al., 2003) because the process of creating a false story (i.e. about something that did not happen) consumes additional cognitive resources (Richards & Gross, 1999, 2000). When compared to those of truth tellers, deceivers' messages show lower cognitive complexity (Newman et al., 2003). Thus, these deceptive messages have lower average sentence length and lower average word length (Zhou &

---

[3] Lexical structure is defined as the structure of lexemes and lexicons (Huang, 1997). 'Lexeme' refers to the combination of words that conveys a specific meaning which cannot be expressed by the individual words alone. 'Lexicon' refers to the vocabulary of the language.

Zhang, 2008). Deceptive communication is also associated with the use of more motion verbs (e.g. *walk*, *move*, *go*), as they provide simpler and more concrete descriptions (and are thus easier to piece together) than words that focus on evaluations and judgments (e.g. *think*, *believe*) (Newman et al., 2003).

3. Emotion

Several studies have shown that emotionally arousing stories tend to attract audience selection and exposure (Knobloch-Westerwick, 2015; Kim, 2015; Lewandowsky, Ecker, Seifert, Schwarz, & Cook, 2012; Berger, 2011; Zillmann, Chen, Knobloch, & Callison, 2004). In addition, recent empirical evidence indicates that emotionally evocative content is more 'viral' than neutral content[4] (Berger and Milkman, 2012; Berger, 2014; Kim, 2015). In particular, online content that evokes high-arousal emotions such as anger and anxiety has been shown to spread faster and more broadly than neutral content (Cotter, 2008; Peters, Kashima, & Clark, 2009; Berger & Milkman, 2012; "Most influential emotions on social networks revealed," 2013). Nonetheless, online content related to negative emotion tends to fade more rapidly (i.e. they are less persistent over time) than content related to positive emotion (Wu, Tan, Kleinberg, & Macy, 2011).

While all three dimensions independently feature in other frameworks, the existing measures for these dimensions are either too subjective or defy easy use by lay readers. In addition, these frameworks tend to focus on the validation of news sources rather than on the content of the fake news itself – a major drawback when it comes to news with limited reference sources. We propose a framework that measures the three dimensions through the following metrics:

1. Lexical structure (as measured through the percentage or proportion of specific classes of words).
2. Level of simplicity (as measured through the readability index).
3. Intensity of emotion in terms of valence.

On the basis of these three dimensions, we create the LeSiE framework (See Figure 1) for identifying fake news:

---

[4] According to Rime (2009), people engage in the sharing of emotionally arousing content because it confers both intrapersonal (e.g. sense-making of the emotional experience) and interpersonal (e.g. establishment or strengthening of social bonds) benefits.

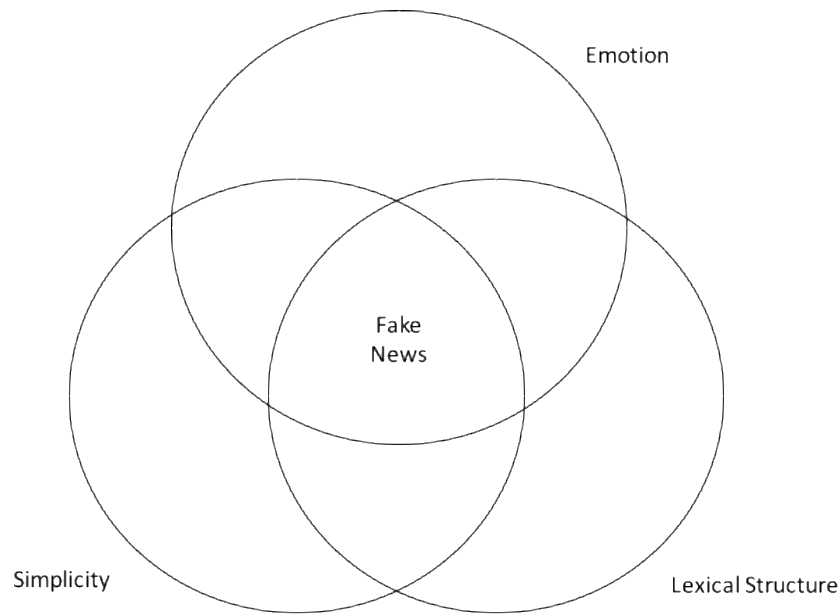

**Fig 1. LeSiE Framework**

**Validation of the Framework**

To validate the framework, we tested it on real data. There are several sets of online data for fake news detection. To facilitate our research, we extracted headlines from www.politifact.com, which contains different types of ratings for different news items. The site classifies news into six different categories, ranging from 'blatant lies' (marked as 'pants on fire') to 'truths'. The different levels of classification are useful, as they present a spectrum of misinformation rather than a binary true/false scenario. If there are lexical structures, levels of simplification, and types of emotion that are unique to fake news, then the degree of such characteristics should (in theory) be different across this spectrum.

Objective measurements of lexical structure, simplification and emotion are critical to the study. These measurements should also be easy for lay people to compute via simple formulae or instructions. We propose measuring emotions through the use of sentiment dictionaries; measuring simplification through the application of readability formulae; and assessing lexical structures by identifying the proportion of classes of words.

**Emotion/sentiment analysis**: Detecting emotion and measuring sentiment polarity have been focal areas in social media analysis, online forum studies, and natural language processing. While there are many approaches to sentiment analysis, the most common involves the application of dictionaries of keywords. In this study, we used Henry's dictionary (Henry, 2008) and the Loughran-McDonald dictionary (Loughran & McDonald, 2011), which are part of the sentimentAnalysis software.

**Readability metrics**: Readability metrics are typically based on features such as the average number of words or characters per sentence, the log of the common usage frequency of the words in the sentences, and selected syntactic and grammatical features. The most famous of these metrics are Flesch-Kincaid (Kincaid, Fishburne, Rogers, & Chissom, 1975), Gunning Fog Index (Gunning, 1952), Coleman-Liau (Coleman & Liau, 1975), SMOG (McLaughlin, 1969) and Automated Readability Index (Smith & Senter, 1967). The Coleman-Liau index (Coleman & Liau, 1975) was used for this study:

CLI = 0.0588 L - 0.296S -15.8.

where L is the average number of letters per 100 words, and S is the number of sentences. As most headlines do not have sentences, the formulae can be reduced to:

CLI = 0.0588L + X

where X is a constant. Given that headlines are less than 100 words, the whole formulae can be simplified to the number of letters per word in the headline.

**Lexical structure**: To assess lexical structures, we used the Stanford NLP parser to parse headlines and determine the proportion of verbs, subjects, and nouns (amongst other classes of words) as described in Chen et. al (2015).

To validate the framework, we tested three hypotheses.
1. There are differences between categories of truth in terms of sentiment.
2. There are differences between categories of truth in terms of readability.
3. There are differences between categories of truth in terms of lexical structure.

We used Tukey's test to evaluate differences between the categories in terms of sentiment, readability and lexical structure. Tukey's test is a multiple comparison test that determines whether the means of different categories are statistically significant from one another.

There are three main assumptions in Tukey's test:
1. The observations being tested are independent within and among the groups.
2. The groups associated with each mean in the test are normally distributed.
3. There is equal within-group variance across the groups associated with each mean in the test.

A variant of the t-test, Tukey's test corrects for family-wise error when there are multiple comparisons being made for the same data (Tukey, 1949). This feature makes it a more suitable candidate than the t-test for multiple comparisons of ranges.

The formula for Tukey's test is:

$$S_q = \frac{Y_A - Y_B}{SE}$$

where $Y_A$ is the larger of the two means being compared, $Y_B$ is the smaller of the two means being compared, and SE is the standard error of the sum of the means. If the $S_q$ value is larger than the critical value $S_a$ obtained from studentized range distribution, the two means are said to be significantly different at level $a, 0 \leq a \leq 1$. The null hypothesis for Tukey's test assumes that all means compared are from the same population (i.e. $\mu_1 = \mu_2 = \mu_3 = ... = \mu_k$). Using this null hypothesis as the template, we formalize the hypotheses to validate the framework as follows:

1. $H_0$: $Sentiment_{Pants\text{-}on\text{-}fire}$ = $Sentiment_{False}$ = $Sentiment_{Barely\ True}$ = ... = $Sentiment_{True}$
2. $H_0$: $Readability_{Pants\text{-}on\text{-}fire}$ = $Readability_{False}$ = $Readability_{Barely\ True}$ = ... = $Readability_{True}$
3. $H_0$: $Lexical_{Pants\text{-}on\text{-}fire}$ = $Lexical_{False}$ = $Lexical_{Barely\ True}$ = ... = $Lexical_{True}$

Unlike most hypothesis tests, we are not interested in a simple accept/reject scenario. In this validation, we will compute the $a$ level across the pairwise comparison. After which, we will consider how the different categories differ from one another at different levels of significance. We expect different categories of truth to show some form of similarity or dissimilarity. We expect the 'truth' and 'mostly true' categories to be different from the other categories. In addition, we expect 'pants on fire' to be a class of its own on some measures.

**Experimentation Results**

Our harvested data contained 11,523 headlines. There were 955 'pants-on-fire' headlines, 2,257 'false' headlines, 1,891 'barely true' headlines, 2,362 'half true' headlines, 2,213 'mostly true' headlines, and 1,845 'true' headlines. Each of the headline categories is almost equally distributed, except for 'pants-on-fire' headlines. During the testing process, we did log transformation to ensure normal distribution for the values.

**Emotions**: We apply the sentiment calculation algorithm as defined in the LM and QDAP dictionaries in R. The results are shown in table 1.

| SA-LM | Pants on Fire | False | Barely True | Half True | Mostly True | True |
|---|---|---|---|---|---|---|
| Pants on Fire | 1.000 | 0.742 | 0.998 | 0.975 | 0.156 | 0.002 |
| False | 0.742 | 1.000 | 0.864 | 0.963 | 0.767 | 0.024 |
| Barely True | 0.998 | 0.864 | 1.000 | 0.999 | 0.153 | 0.001 |
| Half True | 0.975 | 0.963 | 0.999 | 1.000 | 0.253 | 0.001 |
| Mostly True | 0.156 | 0.767 | 0.153 | 0.253 | 1.000 | 0.454 |
| True | 0.002 | 0.024 | 0.001 | 0.001 | 0.454 | 1.000 |

**Table 1: Tukey Test for sentiment value defined by SA LM algorithm implemented in R**

In the case of LM dictionary, the 'mostly true' and 'true' categories are distinct from the other categories. The sentiment values for the 'true' category are statistically different from the other categories, except for the 'mostly true' category. This implies that the 'true' category is different from the other categories in terms of its sentiment value.

| SA QDAP | Pants on Fire | False | Barely True | Half True | Mostly True | True |
|---|---|---|---|---|---|---|
| Pants on Fire | 1.000 | 0.343 | 0.308 | 0.212 | 0.506 | 0.144 |
| False | 0.343 | 1.000 | 1.000 | 1.000 | 0.999 | 0.988 |
| Barely True | 0.308 | 1.000 | 1.000 | 1.000 | 0.997 | 0.997 |
| Half True | 0.212 | 1.000 | 1.000 | 1.000 | 0.987 | 0.999 |
| Mostly True | 0.506 | 0.999 | 0.997 | 0.987 | 1.000 | 0.935 |
| True | 0.144 | 0.988 | 0.997 | 0.999 | 0.935 | 1.000 |

**Table 2: Tukey Test for sentiment value defined by SA QDAP algorithm implemented in R**

In the case of QDAP dictionary, we observe no statistically significant distinction between the categories up to the $a$ level of 0.10. However, the 'pants on fire' category is slightly different from the rest depending on the $a$ level. The other categories do not display the same effect.

**Simplicity**: We applied the Coleman-Liau algorithm as defined in R. The results are shown in Table 3:

| Coleman-Liau | Pants on Fire | False | Barely True | Half True | Mostly True | True |
|---|---|---|---|---|---|---|
| Pants on Fire | 1.000 | 0.997 | 0.967 | 0.895 | 0.024 | 0.017 |
| False | 0.997 | 1.000 | 0.998 | 0.970 | 0.008 | 0.006 |
| Barely True | 0.967 | 0.998 | 1.000 | 1.000 | 0.055 | 0.039 |
| Half True | 0.895 | 0.970 | 1.000 | 1.000 | 0.080 | 0.058 |
| Mostly True | 0.024 | 0.008 | 0.055 | 0.080 | 1.000 | 1.000 |
| True | 0.017 | 0.006 | 0.039 | 0.058 | 1.000 | 1.000 |

**Table 3: Tukey Test for Coleman Liau index implemented in R**

The results from the Coleman-Liau algorithm are similar to the LM dictionary results. The differences between the 'Mostly True'/'True' categories and the other categories are statistically significant at the $a$ level of 0.10. This neatly separates the six categories into two groups (i.e. 'Mostly True'/True' vs. the rest).

**Lexical Structure**: To analyse lexical structure, we applied the Stanford NLP algorithm as defined in R to calculate the proportion of certain classes of words. The results are shown in tables 4-8:

| Adjective | Pants on Fire | False | Barely True | Half True | Mostly True | True |
|---|---|---|---|---|---|---|
| Pants on Fire | 1.000 | 0.214 | 0.005 | 0.000 | 0.000 | 0.000 |
| False | 0.214 | 1.000 | 0.502 | 0.007 | 0.000 | 0.000 |
| Barely True | 0.005 | 0.502 | 1.000 | 0.626 | 0.000 | 0.032 |
| Half True | 0.000 | 0.007 | 0.626 | 1.000 | 0.049 | 0.589 |
| Mostly True | 0.000 | 0.000 | 0.000 | 0.049 | 1.000 | 0.886 |
| True | 0.000 | 0.000 | 0.032 | 0.589 | 0.886 | 1.000 |

**Table 4: Tukey Test for proportion of adjectives identified by Korpus package in R**

| Modal | Pants on Fire | False | Barely True | Half True | Mostly True | True |
|---|---|---|---|---|---|---|
| Pants on Fire | 1.000 | 0.969 | 1.000 | 0.705 | 0.081 | 0.013 |
| False | 0.969 | 1.000 | 0.992 | 0.962 | 0.144 | 0.017 |

| Modal | Pants on Fire | False | Barely True | Half True | Mostly True | True |
|---|---|---|---|---|---|---|
| Barely True | | 1.000 | 0.992 | 1.000 | 0.740 | 0.043 | 0.004 |
| Half True | | 0.705 | 0.962 | 0.740 | 1.000 | 0.573 | 0.138 |
| Mostly True | | 0.081 | 0.144 | 0.043 | 0.573 | 1.000 | 0.951 |
| True | | 0.013 | 0.017 | 0.004 | 0.138 | 0.951 | 1.000 |

**Table 5: Tukey Test for proportion of modal identified by Korpus package in R**

| Name | Pants on Fire | False | Barely True | Half True | Mostly True | True |
|---|---|---|---|---|---|---|
| Pants on Fire | 1.000 | 0.000 | 0.000 | 0.000 | 0.000 | 0.000 |
| False | 0.000 | 1.000 | 0.502 | 0.000 | 0.000 | 0.000 |
| Barely True | 0.000 | 0.502 | 1.000 | 0.005 | 0.000 | 0.000 |
| Half True | 0.000 | 0.000 | 0.005 | 1.000 | 0.000 | 0.098 |
| Mostly True | 0.000 | 0.000 | 0.000 | 0.000 | 1.000 | 0.625 |
| True | 0.000 | 0.000 | 0.000 | 0.098 | 0.625 | 1.000 |

**Table 6: Tukey Test for proportion of names identified by Korpus package in R**

| Number | Pants on Fire | False | Barely True | Half True | Mostly True | True |
|---|---|---|---|---|---|---|
| Pants on Fire | 1.000 | 0.476 | 0.002 | 0.000 | 0.000 | 0.000 |
| False | 0.476 | 1.000 | 0.089 | 0.000 | 0.000 | 0.000 |
| Barely True | 0.002 | 0.089 | 1.000 | 0.003 | 0.000 | 0.000 |
| Half True | 0.000 | 0.000 | 0.003 | 1.000 | 0.002 | 0.041 |
| Mostly True | 0.000 | 0.000 | 0.000 | 0.002 | 1.000 | 0.984 |
| True | 0.000 | 0.000 | 0.000 | 0.041 | 0.984 | 1.000 |

**Table 7: Tukey Test for proportion of numbers identified by Korpus package in R**

| Verb | Pants on Fire | False | Barely True | Half True | Mostly True | True |
|---|---|---|---|---|---|---|
| Pants on Fire | 1.000 | 0.089 | 0.549 | 0.000 | 0.000 | 0.000 |
| False | 0.089 | 1.000 | 0.868 | 0.008 | 0.000 | 0.000 |
| Barely True | 0.549 | 0.868 | 1.000 | 0.000 | 0.000 | 0.000 |
| Half True | 0.000 | 0.008 | 0.000 | 1.000 | 0.099 | 0.764 |
| Mostly True | 0.000 | 0.000 | 0.000 | 0.099 | 1.000 | 0.875 |
| True | 0.000 | 0.000 | 0.000 | 0.764 | 0.875 | 1.000 |

**Table 8: Tukey Test for proportion of verbs identified by Korpus package in R**

Across the classes of word identified, we observed statistically significant distinctions between the categories up to the $a$ level of 0.10. Profound and distinct differences are found for the verb, number, name and adjective classes. The verb class separated the six categories into two groups – those with more truth versus those with less truth. Separations for the adjective, name and number classes can also be seen across almost all the categories.

**Conclusion**

Audiences often do not possess the necessary literacy skills to interpret news critically (Hango, 2014). Given the serious nature and proliferation of fake news, it is critical to create a simple diagnostic framework that can be used by lay people. We developed a three-dimensional framework (LeSiE) that simplifies the identification of fake news by doing away with the need for calculators or complex calculations. Our tests of the framework show that the following cues and hand calculations can point to the potential presence of fake news:

1. Strong positive or negative words.
2. Length of the title in terms of number of letters.
3. Preponderance of verbs, adjectives, names or numbers.

The task of seeing through misinformation just got easier.